\documentclass[aps,pra,preprint,groupedaddress,showpacs,%
tightenlines,
nofootinbib,floatfix]{revtex4}

\usepackage{amsmath,amssymb,amsfonts,bm}
\usepackage{graphicx}
\usepackage{dcolumn}
\usepackage{mathrsfs}

\begin{document}

\title{Shift and broadening of resonance lines of antiprotonic helium
  atoms in solid helium}

\author{Andrzej Adamczak}
\email{andrzej.adamczak@ifj.edu.pl}
\affiliation{Institute of Nuclear Physics, Polish Academy of Sciences,
Radzikowskiego 152, PL-31342~Krak\'ow, Poland}

\author{Dimitar Bakalov}
\email{dbakalov@inrne.bas.bg}
\affiliation{Institute for Nuclear Research and Nuclear Energy,
  Bulgarian Academy of Sciences, Tsarigradsko chauss\'ee 72,
  Sofia 1784, Bulgaria}

\date{\today}

\begin{abstract}
  We have estimated the shift and broadening of the resonance lines in
  the spectrum of antiprotonic helium atoms $\bar{p}\mathrm{He}^{+}$
  implanted in solid helium $^4$He. The application of the response
  function for crystalline helium has enabled determination of the
  contributions from the collective degrees of freedom to the shift and
  broadening. It occurs that the broadening due to the collective
  motion is negligible compared to the natural line width. The
  available pair-correlation functions for crystalline $^4$He have been
  applied for estimating the resonance-line shift due to collisions of
  $\bar{p}\mathrm{He}^{+}$ atom with the surrounding $^4$He atoms. The
  dependence of the line shift, which has been calculated in the
  quasistatic limit, on the solid-$^4$He density is nonlinear.
\end{abstract}


\maketitle

\section{Introduction}
\label{sec:intro}

The aim of this work is estimation of the shift and broadening of
resonance lines of the $\bar{p}\mathrm{He}^{+}$ atoms in solid $^4$He as
functions of helium density. A similar study for liquid $^4$He,
presented in Ref.~\cite{adam13}, showed that the shift is a linear
function of density in normal-fluid $^4$He and displays only small
oscillations with temperature in superfluid $^4$He. The
pair-distribution function, which describes the atom distribution around
a given particle, is very similar for both the normal-fluid and the
superfluid helium at various temperatures and saturated-vapor pressure.
Also, the density change for liquid $^4$He at such a pressure is lower
than about 15\%.  Therefore, solid $^4$He gives the possibility of
studying the resonance-line shift and broadening at much higher
densities and thus for shorter distances between the atoms.

The antiprotonic helium atoms are created when antiprotons are
decelerated in helium targets and then replace one electron in the
helium atoms. About 3\% of antiprotons are captured in metastable states
$(n,\ell)$ with long lifetimes on the order of
microseconds~\cite{yamaz02}. This phenomenon enabled high-precision
laser spectroscopy of $\bar{p}\mathrm{He}^{+}$ atoms. As a result, the
antiproton-to-electron mass ratio~\cite{hori11} has been determined with
the best accuracy up-to-date. Such high-precision measurements required
the estimation of different systematic effects. Among the most important
effects are the shift and broadening of the spectral lines due to
interaction with the helium atoms and varying with the density of the
helium target. In the case of helium gas, these effects have been
calculated in the semiclassical approach with the use of a pairwise
potential of interaction between the $\bar{p}\mathrm{He}^{+}$ atom and
an ordinary helium atom~\cite{baka00}. The calculation results agree
well with the experimental data up to the gaseous-helium density
$\varrho=127$~g/l \cite{yamaz02}. It has been found that the resonance
line shifts in these gas targets are proportional to the helium density
within the experimental accuracy~\cite{tori99,hori06}.  The attempts for
laser spectroscopy of antiprotonic helium atoms in liquid helium, which
were performed by the ASACUSA Collaboration at CERN~\cite{hori-private},
encouraged us to evaluate the corresponding shifts and broadening in
fluid and superfluid $^4$He~\cite{adam13}. In the case of liquid helium,
it was necessary to take into account the influence of collective
dynamics of helium atoms on the shifts and broadening of spectral lines,
apart from the collisional effects. Our calculations were performed for
the target temperatures $T=1.0$--4.27~K at pressures~$P$ up to 8~bars,
which corresponds to the maximum pressure applied in the gaseous-helium
experiments. For such conditions, the maximum density of about 146~g/l
is reached at the $\lambda$~point. Our calculations confirmed the linear
dependence of the resonance-line shift in normal-fluid $^4$He. On the
other hand, an appreciable oscillation (9\%) of this shift as a~function
of temperature was found for the region of superfluid, where the $^4$He
density is practically constant.

At the atmospheric pressure, helium remains liquid even at absolute
zero, due to the weak interaction between the helium atoms and the large
zero-point motion of these atoms. However, the application of pressure
$P\approx{}25$~bar below a few kelvins leads to solidification of
$^4$He~\cite{glyd87}. The solid helium at the melting line has a density
significantly greater than that of liquid helium. The solid-$^4$He
density can be increased using even larger pressures.

In this work, the numerical calculations are performed for the
transition $|i\rangle=(n,\ell)=(39,35)\to|f\rangle=(n',\ell')=(38,34)$
between the initial $|i\rangle$ and final $|f\rangle$ states of the
$\bar{p}\mathrm{He}^{+}$ atom (transition~1), which has been
experimentally observed even at a~relatively high density of
127~g/l~\cite{hori11}. The resonance wavelength for this transition
equals $\lambda_0=5972.570$~\AA~\cite{tori99} and the corresponding
resonance energy is $E_0=2.07589$~eV. In the case of this line, the
Auger decay rate $R_A\approx{}1.11\times{}10^8$~s$^{-1}$~\cite{yamag02}
determines the natural line width $\Gamma_0$, according to the relation
\begin{equation}
  \label{eq:gam_Auger}
  \Gamma_0 \approx \hbar R_A  \,,
\end{equation}
which gives $\Gamma_0\approx{}0.73\times{}10^{-7}$~eV. The corresponding
frequency $\nu_n=\Gamma_0/h$ equals 0.018~GHz.

In Sec.~\ref{sec:solid_he} we estimate the changes of the line profile
due to the collective motion in the quantum crystal $^4$He, using the
method presented in Ref.~\cite{sing60}. The line shift, which is caused
by collisions of $\bar{p}\mathrm{He}^{+}$ atom with neighboring $^4$He
atoms, is calculated in Sec.~\ref{sec:single-atom} in the quasistatic
approximation of Ref.~\cite{adam13} using the available pair-correlation
functions $g(r)$ for crystalline $^4$He.  It has been shown
in~Ref.~\cite{adam13} that in the quasistatic limit of slow collisions
the semiclassical expression for the collisional shift of
Ref.~\cite{baka00} takes the form of a mean value of the interatomic
potential, averaged over the spatial distribution of the perturbing
helium atoms around the emitting antiprotonic atoms. Under the
assumption that the latter is close to the distribution in pure helium,
described by the pair-correlation function, this allows for using
experimental data about $g(r)$.  Unfortunately the semiclassical
expression for the line broadening \cite{baka00} does not take any
simple form in the quasistatic limit, so that the collisional width
remains to be evaluated by full scale semiclassical or quantum
calculations, which is beyond the scope of the present Brief Report. The
results are briefly discussed in Sec.~\ref{sec:concl}.

\section{Line shift and broadening due to the collective dynamics of
  solid helium}
\label{sec:solid_he}

The shift and broadening of a~resonance line can be evaluated using the
method developed by Singwi and Sj\"olander~\cite{sing60}, which employs
the Van Hove formalism of the response function~\cite{vanh54}. When a
particle which absorbs or emits a photon is set at a fixed position, the
absorption cross section $\sigma_a$ takes the following form
\begin{equation}
  \label{eq:sig_abs_fixed}
  \sigma_a(E)=\sigma_0\, \frac{\Gamma_0^2/4}{(E-E_0)^2+\Gamma_0^2/4} \,
\end{equation}
where $E$ is the photon energy, $\sigma_0$ is the resonance maximum at
the resonance energy~$E_0$ and $\Gamma_0$ stands for the natural width
of the resonance. In the case of a harmonic crystal, the resonance
profile $\sigma_a^\textrm{solid}(E)$ can be rigorously derived. For a
monoatomic cubic Bravais lattice, the exact form of the profile is given
as~\cite{sing60}
\begin{equation}
  \label{eq:sig_abs_sol}
  \begin{split}
    \sigma_a^\textrm{solid}(E) = \frac{\pi\sigma_0\Gamma_0}{2}\,
    & \exp(-2W) \biggl[ \frac{1}{2\pi}\,
    \frac{\Gamma_0}{(E-E'_0)^2+\tfrac{1}{4}\Gamma_0^2} \\
    & +\sum_{n=1}^{\infty} g_n(\omega,T)\, \frac{(2W)^n}{n!} \biggr] ,
  \end{split}
\end{equation}
where $\hbar\omega$ and $\hbar{}q$ denote the energy and momentum
transfer to the crystal, respectively, and $T$ is temperature. Although
solid $^4$He has the hcp structure under specific conditions~(see the
phase diagram, e.g., in Ref.~\cite{glyd87}), apart from the cubic bcc
and fcc structures, the above expansion establishes a fair approximation
also for this lattice. In the case of laser-stimulated transitions in
the antiprotonic helium, the resonance energy in
Eq.~(\ref{eq:sig_abs_sol}) equals $E'_0=E_0+\Delta{}E_0$, where
$\Delta{}E_0$ is the line shift due to the pairwise interaction.  The
exponent $2W$ of the Debye-Waller factor $\exp(-2W)$ can be expressed as
follows
\begin{equation}
  \label{eq:2W}
  2W = E_\mathrm{r} \int_0^{\infty} \frac{Z(w)}{w}
  \coth\left( \tfrac{1}{2} \beta_T w \right) \mathrm{d}w ,
  \quad \beta_T = \frac{1}{k_\mathrm{B}T} ,
\end{equation}
where $Z(w)$ is the normalized density of vibrational states in the
crystal, $k_\mathrm{B}$ is Boltzmann's constant and $E_\mathrm{r}$
denotes the recoil energy
\begin{equation}
  \label{eq:E_r}
  E_\mathrm{r} = \frac{(\hbar q)^2}{2M} \,,
\end{equation}
in which $M$ is the mass of antiprotonic helium.

The first term in the expansion~(\ref{eq:sig_abs_sol}) describes the
recoil-less photon absorption or emission in the rigid crystal lattice.
The next terms of this expansion, which are proportional to $q^{2n}$,
describe the same process with simultaneous absorption or emission of
one or more phonons. The functions~$g_n$ from Eq.~(\ref{eq:sig_abs_sol})
are defined in~Ref.~\cite{sing60}. In particular, the one-phonon term
$2Wg_1$ in the brackets of this equation takes the following form
\begin{equation}
  \label{eq:g1}
  2W  g_1(\omega,T) =
  E_r \frac{Z(\omega)}{\omega} \, [n_\mathrm{B}(\omega,T)+1] \,,
\end{equation}
where
\begin{equation}
  \label{eq:Bose_fac}
  n_\mathrm{B}(\omega,T) = [\exp(\beta_T\omega)-1]^{-1}
\end{equation}
is the Bose population factor for phonons. The amplitudes of all the
processes are determined by the Debye-Waller factor. When $2W\ll{}1$ for
a~specific photon energy, target and temperature (strong-binding case),
the recoil-less term is significant. Such a situation takes place in the
case of the M\"ossbauer effect.

When a~photon is absorbed or emitted by the antiprotonic helium atom,
the momentum and energy transfers to the crystal lattice are equal to
\begin{equation}
  \label{eq:transf}
  \hbar q = p \,, \qquad \hbar\omega  = E-E'_0 \,,
\end{equation}
respectively. The absolute value of the photon momentum is denoted here
by~$p$. In the case of transition~1, we have
\begin{equation}
  \label{eq:q_value}
  q = 2\pi / \lambda_0 = 0.001052\, \textrm{\AA}^{-1}
\end{equation}
and the recoil energy equals $E_r=0.461764\times 10^{-9}$~eV. Thus, the
recoil energy is very small compared to the resonance energy
$E_r/E_0\approx{}2.2\times{}10^{-10}$.

The Debye-Waller factor can be estimated using the Debye model of
isotropic crystal, which is a~fair approximation also for quantum
crystals such as solid helium, hydrogen or deuterium. In this model, the
density of vibrational states takes the form
\begin{equation}
  \label{eq:Z_Debye}
  Z(w)=
  \begin{cases}
    3\, w^2/w_\mathrm{D}^3 & \mathrm{~if~} \,
    w \leq w_\mathrm{D} \\
    0 & \text{~if~} \, w > w_\mathrm{D} \,,
  \end{cases}
\end{equation}
in which the maximum energy of vibrations $w_\mathrm{D}$ is determined
by the Debye temperature~$\theta_\mathrm{D}$ of the crystal:
$w_\mathrm{D}=k_\mathrm{B}\theta_\mathrm{D}$. The Debye temperature for
solid $^4$He is greater than 25~K. For the pressures 26.7--129~bar and
the corresponding temperatures 1~K--4~K, which are considered in this
work, $\theta_\mathrm{D}\approx{}25$~K--38~K \cite{tric72} and thus
$T/\theta_\mathrm{D}\ll{}1$. In the limit $T\to{}0$, Eq.~(\ref{eq:2W})
is expressed by a~simple integral. As a result, we obtain the following
expression:
\begin{equation}
  \label{eq:2W_approx}
  2W = \frac{3}{2} \frac{E_\mathrm{r}}{w_\mathrm{D}} \,,
\end{equation}
which is a good approximation for $T/\theta_\mathrm{D}\ll{}1$. In the
case of transition~1, one has $2W\sim{}10^{-7}$. Thus, the recoil-less
term in the expansion~(\ref{eq:sig_abs_sol}) is dominant and the
subsequent phonon contributions are negligible. This means that the
resonance-line shift in solid helium is solely due to the collisional
correction $\Delta{}E_0$. Let us note that in the case of a free
$\bar{p}\mathrm{He}^{+}$ atom the line shift is strictly equal to the
recoil energy~$E_r$. On the other hand, in solid helium, the recoil
effect disappears since the response of the $^4$He lattice to the
resonance-photon absorption or emission is practically the response of a
rigid lattice.

The phonon contribution to the line broadening is determined by the
one-phonon term~(\ref{eq:g1}) with the width determined by
$w_\mathrm{D}\approx{}2$--3~meV. However, this contribution can be
neglected because of an extremely small amplitude of the phonon
processes, which is caused by smallness of the $2W$ factor. Therefore,
the collective degrees of freedom practically do not change the
resonance-line width in solid $^4$He. The width of the dominant
recoil-less term is equal to the natural resonance width~$\Gamma_0$.

\section{Collisional shift of resonance lines in solid helium}
\label{sec:single-atom}

The collisional shift $\Delta{}E_0$ of the resonance lines in
crystalline $^4$He is estimated here in the quasistatic limit using the
method that has been discussed in detail in Ref.~\cite{adam13}. The line
shift is expressed in terms of the pairwise potentials of
$\bar{p}\mathrm{He}^{+}$ interaction with a single helium atom and the
pair-correlation function $g(r)$ of a condensed $^4$He target for a
fixed temperature and pressure. The expression $g(r)\,\mathrm{d}r$
gives the probability of finding a $^4$He atom in the shell
$[r,r+\mathrm{d}r]$ around a given atom.  We use here the spherically
symmetric pairwise potentials $V_i(r)$ and $V_f(r)$ of the
$\bar{p}\mathrm{He}^{+}$-He interaction in the initial and final states
of antiprotonic helium~\cite{baka00,baka12}. As a result, the
collisional contribution $\Delta{}E_0$ to the resonance-line shift is
approximated by the following expression:
\begin{equation}
  \label{qst}
  \Delta E_0=\int_0^{r_\mathrm{max}} \mathrm{d} r\, g(r) \Delta V(r) \,,
\end{equation}
where $\Delta{}V(r)=V_f(r)-V_i(r)$ and $r_\mathrm{max}$ is a cutoff. Let
us note that the radius~$r$ in the function $g(r)$ denotes the distance
reckoned from a~given $^4$He atom located in the origin. In our case, we
replace this atom by the implanted antiprotonic helium atom. However,
this is a reasonable approximation since the probability density
calculated for the two-particle system $\bar{p}\mathrm{He}^{+}$+He is
very similar to $g(r)$ at $r\lesssim{}3$~\AA~\cite{adam13}.

\begin{figure}[htb]
  \centering
  \includegraphics[width=8cm]{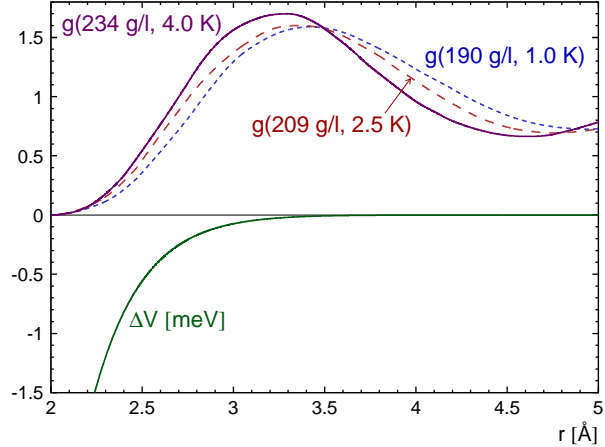}
  \caption{(Color online) The difference $\Delta{}V(r)$ of the pairwise
    potentials $V_{(39,35)}(r)$ and $V_{(38,34)}(r)$ together with the
    pair-correlation functions~$g(r)$ for solid $^4$He
    \cite{clar08,whit79} versus the distance~$r$ between the
    antiprotonic helium and the $^4$He atom. The densities~$\varrho$ and
    and temperatures~$T$ of the three targets are given in the plot.
    One can conclude from this figure that the contribution to the
    resonance energy shift, Eq.~(\ref{qst}), practically comes from the
    interval 2.0~\AA~$<r{}\lesssim$~3.3~\AA, where both $\Delta{}V(r)$
    and $g(r)$ have significant values. For $r\leq{}2.0$~\AA, the
    correlation function disappears due to the finite size and the
    short-distance repulsion of helium atoms.}
    \label{fig:gpot}
\end{figure}

In the literature, the data regarding the function $g(r)$ are scarce. We
use here the theoretical $g(r)$ for the solid $^4$He near the melting
curve at temperature $T=1.0$~K and pressure $P=26.7$~bar
($\rho=190$~g/l)~\cite{clar08}. Also, we employ the theoretical $g(r)$
for $T=2.5$~K at $P=57$~bar ($\rho=209$~g/l) and $T=4.0$~K at
$P=129$~bar ($\rho=234$~g/l)~\cite{whit79}.
%
\begin{figure}[htb]
  \centering
  \includegraphics[width=8cm]{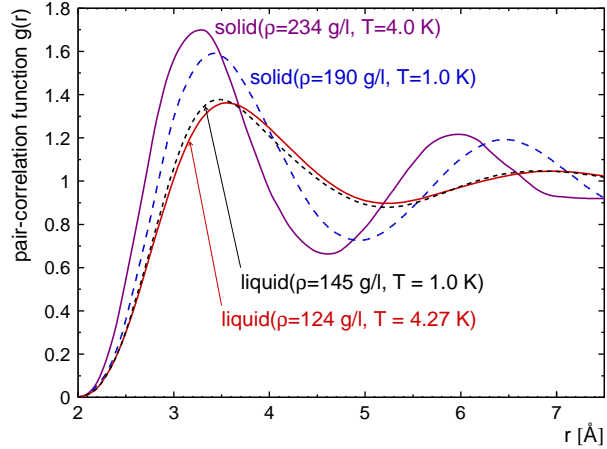}
  \caption{(Color online) The pair-correlation functions $g(r)$ for
    solid~\cite{clar08,whit79} and liquid (at the saturated-vapor
    pressure)~\cite{sear79} $^4$He at several values of target
    density~$\varrho$ and temperature~$T$.}
    \label{fig:g_sol_liq}
\end{figure}%
The functions $g(r)$ for solid and liquid $^4$He are compared in
Fig.~\ref{fig:g_sol_liq} for $T=1$~K and $T\approx{}4$~K. Although the
liquid $^4$He is superfluid at 1~K and normal fluid at 4.27~K, the
corresponding pair-correlation functions are very similar. Therefore, no
significant change of behavior of the resonance-line shift is expected
in liquid helium. On the other hand, the functions $g(r)$ for solid
$^4$He at the presented temperatures and densities differ significantly
from each other and from the corresponding functions for liquid helium.
In particular, one can see that the probability of finding a neighboring
$^4$He atom at 2~\AA~$<r<3.5$~\AA\ is much greater in the solid targets.
This interval of~$r$ gives a dominant contribution to the line shift,
which is shown in Fig.~\ref{fig:gpot}. As a result, one can expect a
significant change of the line-shift behavior in solid helium.
%

\begin{figure}[htb]
  \centering
  \includegraphics[width=8cm]{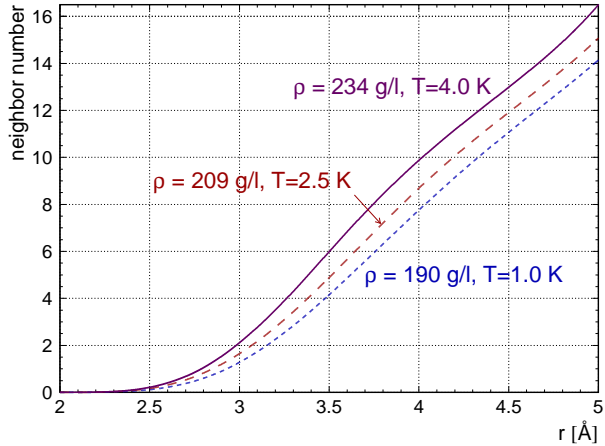}
  \caption{(Color online) The number $n(r)$ of $^4$He atoms within the
    sphere of radius~$r$ that surround an atom placed at $r=0$ in solid
    helium.}
  \label{fig:neighbors_solid}
\end{figure}

The average number $n(r)$ of $^4$He atoms,
\begin{equation}
  \label{eq:neighbors}
  n(r) = 4\pi N_0 \int_0^r \mathrm{d}r'\, r'^2 g(r') \,,
\end{equation}
which are located within the sphere of radius~$r$ around the helium atom
in the origin, is shown in~Fig.~\ref{fig:neighbors_solid} for the three
pressures.
Since $n(r)<2$ for $r\lesssim{}3$~\AA, where the absolute value of
$\Delta{}V(r)$ has the largest amplitude, the contribution to the
resonance-line shift~(\ref{qst}) from this region is dominant.
Therefore, using the pairwise interaction potentials for determination
of the resonance shift in solid $^4$He is a~reasonable approximation.

The results of our calculations for solid $^4$He for the three different
target densities are summarized in Table~\ref{table:shift},
\begin{table}[htb]
  \begin{center}
    \caption{The resonance-line shift $\Delta{}E_0$ and the reduced line
      shift $\Delta{}E_0/\varrho$ for solid $^4$He.}
    \label{table:shift}
    \begin{ruledtabular}
    \newcolumntype{.}{D{.}{.}{3.3}}
    \begin{tabular}{. . c . .}
      \multicolumn{1}{c}{Temperature}&
      \multicolumn{1}{c}{Pressure}&
      \multicolumn{1}{c}{Density}&
      \multicolumn{1}{c}{$\Delta{}E_0$~~~~}&
      \multicolumn{1}{c}{$\Delta{}E_0/\varrho$}\\
      \multicolumn{1}{c}{[K]}&
      \multicolumn{1}{c}{[bar]}&
      \multicolumn{1}{c}{[g/l]}&
      \multicolumn{1}{c}{[GHz]~~~~}&
      \multicolumn{1}{c}{[GHz~l/g]}\\
      \hline
      1.0 &  26.7 & 190 &  -99.8 & -0.525 \\
      2.5 &  57.0 & 209 & -130.9 & -0.626 \\
      4.0 & 129.0 & 234 & -164.8 & -0.704 \\
  \end{tabular}
  \end{ruledtabular}
  \end{center}
\end{table}%
where the reduced line shift $\Delta{}E_0/\varrho$ is given in the fifth
column. A~dependence of the calculated reduced line shift on the upper
limit $r_\textrm{max}$ in the integral~(\ref{qst}) is shown
in~Fig.~\ref{fig:redshift_solid}.
\begin{figure}[htb]
  \centering
  \includegraphics[width=8cm]{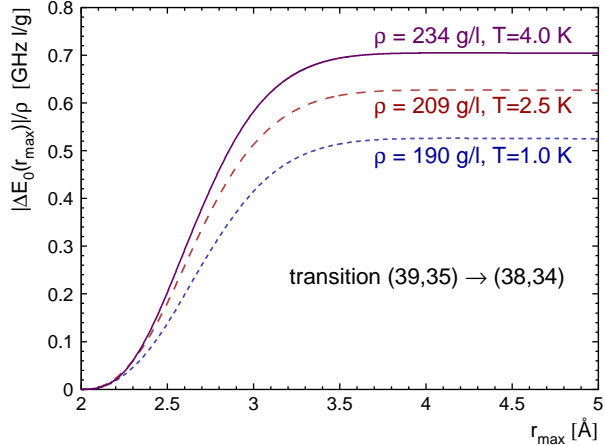}
  \caption{(Color online) The absolute value $|\Delta{}E_0|/\varrho$ of
    the reduced resonance-line shift in solid $^4$He as a~function of
    $r_\mathrm{max}$.}
    \label{fig:redshift_solid}
\end{figure}%
One can see that it is sufficient to perform integration in
Eq.~(\ref{qst}) up to $r_\textrm{max}\approx{}3.8$~\AA\@. Thus, the
knowledge of the pair-correlation function above this limit is not
needed. The resonance-line shifts $\Delta{}E_0$ which are presented in
Table~\ref{table:shift} display a clear nonlinear dependence on the
target density, which is in contrast to the behavior of analogous line
shifts in $^4$He gas~\cite{tori99,hori06} and liquid $^4$He above
$T=2.17$~K~\cite{adam13}. The absolute values of the reduced line shifts
in solid helium are greater than the values of the corresponding line
shifts in liquid $^4$He (e.g., $\Delta{}E_0/\varrho=-0.427$~GHz~l/g for
$\varrho=146$~g/l at $T=2.27$~K \cite{adam13}).

\section{Conclusions}
\label{sec:concl}

The experimental and theoretical investigation of the density dependence
of the resonance-line shifts in gaseous helium and the projects of the
ASACUSA Collaboration at CERN for the high-accuracy laser spectroscopy
of $\bar{p}\mathrm{He}^{+}$ atoms in liquid $^4$He has motivated the
present work. In particular, we have studied the influence of collective
degrees of freedom in solid $^4$He on the broadening and shift of the
resonance lines of antiprotonic helium located in this target.

We have found that the resonant absorption or emission of a laser photon
by the $\bar{p}\mathrm{He}^{+}$ atom implanted in solid $^4$He is a
fully recoil-less process which takes place in the rigid lattice and
thus is analogous the M\"ossbauer effect. This is due to a~very small
momentum transfer to the lattice of about 0.001~\AA$^{-1}$ and extremely
small amplitudes of phonon processes, which are simultaneous with the
resonant transition in the antiprotonic helium. Therefore, there is
exactly no contribution to the resonance-line shifts from the collective
motion in solid helium.  On the other hand, the resonance lines in the
case of $\bar{p}\mathrm{He}^{+}$ atom in a dilute helium gas are shifted
by the corresponding recoil energy, which is however very small. The
broadening of resonance-line shifts due to the collective motion in
crystalline $^4$He is determined by the one-phonon processes and equals
about 2-3~meV, which is a typical maximum phonon energy in solid helium.
However, the amplitude of such processes is smaller by many orders of
magnitude than the recoil-less process so that this broadening cannot be
observed in experiments. As a result, the total broadening and shift of
the resonance lines in solid helium are determined by the collisional
effects. The magnitude of the reduced collisional shift
$\Delta{}E_0/\varrho$ for the resonance transition
$(39,35)\to{}(38,34)$, which has been calculated in the quasistatic
approach, ranges from $-0.525$ to $-0.704$~GHz~l/g when the density of
solid $^4$He varies from 190 to 234~g/l. Therefore, the density
dependence of the total line shift is clearly nonlinear, which is in
contrast to the behavior of the analogous line shifts in gaseous and
normal liquid $^4$He.

In order to improve the accuracy of the presented evaluation of the
collisional contribution to the resonance-line shifts it is
indispensable to calculate the potentials of $\bar{p}\mathrm{He}^{+}$
interaction with at least two helium atoms. Also, this would enable a
reliable estimation of the collisional broadening of the spectral lines.
However, a~calculation of the appropriate potentials is much more
complicated than in the case of one neighboring He atom.

\begin{acknowledgments}
  This work has been performed under the framework of collaboration
  between the Bulgarian Academy of Sciences and the Polish Academy of
  Sciences.
\end{acknowledgments}


\bibliography{apsol}

\begin{thebibliography}{16}
\expandafter\ifx\csname natexlab\endcsname\relax\def\natexlab#1{#1}\fi
\expandafter\ifx\csname bibnamefont\endcsname\relax
  \def\bibnamefont#1{#1}\fi
\expandafter\ifx\csname bibfnamefont\endcsname\relax
  \def\bibfnamefont#1{#1}\fi
\expandafter\ifx\csname citenamefont\endcsname\relax
  \def\citenamefont#1{#1}\fi
\expandafter\ifx\csname url\endcsname\relax
  \def\url#1{\texttt{#1}}\fi
\expandafter\ifx\csname urlprefix\endcsname\relax\def\urlprefix{URL }\fi
\providecommand{\bibinfo}[2]{#2}
\providecommand{\eprint}[2][]{\url{#2}}

\bibitem[{\citenamefont{Adamczak and Bakalov}(2013)}]{adam13}
\bibinfo{author}{\bibfnamefont{A.}~\bibnamefont{Adamczak}} \bibnamefont{and}
  \bibinfo{author}{\bibfnamefont{D.}~\bibnamefont{Bakalov}},
  \bibinfo{journal}{Phys.~Rev.~A} \textbf{\bibinfo{volume}{88}},
  \bibinfo{pages}{042505} (\bibinfo{year}{2013}).

\bibitem[{\citenamefont{Yamazaki et~al.}(2002)\citenamefont{Yamazaki, Morita,
  Hayano, Widmann, and Eades}}]{yamaz02}
\bibinfo{author}{\bibfnamefont{T.}~\bibnamefont{Yamazaki}},
  \bibinfo{author}{\bibfnamefont{N.}~\bibnamefont{Morita}},
  \bibinfo{author}{\bibfnamefont{R.}~\bibnamefont{Hayano}},
  \bibinfo{author}{\bibfnamefont{E.}~\bibnamefont{Widmann}}, \bibnamefont{and}
  \bibinfo{author}{\bibfnamefont{J.}~\bibnamefont{Eades}},
  \bibinfo{journal}{Phys.\ Rep.} \textbf{\bibinfo{volume}{366}},
  \bibinfo{pages}{183} (\bibinfo{year}{2002}).

\bibitem[{\citenamefont{Hori et~al.}(2011)\citenamefont{Hori, S\'ot\'er, Barna
  et~al.}}]{hori11}
\bibinfo{author}{\bibfnamefont{M.}~\bibnamefont{Hori}},
  \bibinfo{author}{\bibfnamefont{A.}~\bibnamefont{S\'ot\'er}},
  \bibinfo{author}{\bibfnamefont{D.}~\bibnamefont{Barna}},
  \bibnamefont{et~al.}, \bibinfo{journal}{Nature}
  \textbf{\bibinfo{volume}{475}}, \bibinfo{pages}{484} (\bibinfo{year}{2011}).

\bibitem[{\citenamefont{Bakalov et~al.}(2000)\citenamefont{Bakalov, Jeziorski,
  Korona, Szalewicz, and Tchukova}}]{baka00}
\bibinfo{author}{\bibfnamefont{D.}~\bibnamefont{Bakalov}},
  \bibinfo{author}{\bibfnamefont{B.}~\bibnamefont{Jeziorski}},
  \bibinfo{author}{\bibfnamefont{T.}~\bibnamefont{Korona}},
  \bibinfo{author}{\bibfnamefont{K.}~\bibnamefont{Szalewicz}},
  \bibnamefont{and} \bibinfo{author}{\bibfnamefont{E.}~\bibnamefont{Tchukova}},
  \bibinfo{journal}{Phys.~Rev.\ Lett.} \textbf{\bibinfo{volume}{84}},
  \bibinfo{pages}{2350} (\bibinfo{year}{2000}).

\bibitem[{\citenamefont{Torii et~al.}(1999)\citenamefont{Torii, Hayano, Hori
  et~al.}}]{tori99}
\bibinfo{author}{\bibfnamefont{H.~A.} \bibnamefont{Torii}},
  \bibinfo{author}{\bibfnamefont{R.~S.} \bibnamefont{Hayano}},
  \bibinfo{author}{\bibfnamefont{M.}~\bibnamefont{Hori}}, \bibnamefont{et~al.},
  \bibinfo{journal}{Phys.\ Rev.~A} \textbf{\bibinfo{volume}{59}},
  \bibinfo{pages}{223} (\bibinfo{year}{1999}).

\bibitem[{\citenamefont{Hori et~al.}(2006)\citenamefont{Hori, Dax, Eades
  et~al.}}]{hori06}
\bibinfo{author}{\bibfnamefont{M.}~\bibnamefont{Hori}},
  \bibinfo{author}{\bibfnamefont{A.}~\bibnamefont{Dax}},
  \bibinfo{author}{\bibfnamefont{J.}~\bibnamefont{Eades}},
  \bibnamefont{et~al.}, \bibinfo{journal}{Phys.~Rev.~Lett.}
  \textbf{\bibinfo{volume}{96}}, \bibinfo{eid}{243401} (\bibinfo{year}{2006}).

\bibitem[{\citenamefont{Hori}()}]{hori-private}
\bibinfo{author}{\bibfnamefont{M.}~\bibnamefont{Hori}}, \bibinfo{note}{private
  communication}.

\bibitem[{\citenamefont{Glyde and Svensson}(1987)}]{glyd87}
\bibinfo{author}{\bibfnamefont{H.~R.} \bibnamefont{Glyde}} \bibnamefont{and}
  \bibinfo{author}{\bibfnamefont{E.~C.} \bibnamefont{Svensson}},
  \emph{\bibinfo{title}{Methods of Experimental Physics}}
  (\bibinfo{publisher}{Academic Pres Inc.}, \bibinfo{address}{London},
  \bibinfo{year}{1987}), vol.~\bibinfo{volume}{23}, p. \bibinfo{pages}{303}.

\bibitem[{\citenamefont{Yamaguchi et~al.}(2002)\citenamefont{Yamaguchi,
  Ishikawa, Sakaguchi et~al.}}]{yamag02}
\bibinfo{author}{\bibfnamefont{H.}~\bibnamefont{Yamaguchi}},
  \bibinfo{author}{\bibfnamefont{T.}~\bibnamefont{Ishikawa}},
  \bibinfo{author}{\bibfnamefont{J.}~\bibnamefont{Sakaguchi}},
  \bibnamefont{et~al.}, \bibinfo{journal}{Phys.\ Rev.~A}
  \textbf{\bibinfo{volume}{66}}, \bibinfo{eid}{022504} (\bibinfo{year}{2002}).

\bibitem[{\citenamefont{Singwi and Sj\"olander}(1960)}]{sing60}
\bibinfo{author}{\bibfnamefont{K.~S.} \bibnamefont{Singwi}} \bibnamefont{and}
  \bibinfo{author}{\bibfnamefont{A.}~\bibnamefont{Sj\"olander}},
  \bibinfo{journal}{Phys.~Rev.} \textbf{\bibinfo{volume}{120}},
  \bibinfo{pages}{1093} (\bibinfo{year}{1960}).

\bibitem[{\citenamefont{{Van Hove}}(1954)}]{vanh54}
\bibinfo{author}{\bibfnamefont{L.}~\bibnamefont{{Van Hove}}},
  \bibinfo{journal}{Phys.~Rev.} \textbf{\bibinfo{volume}{95}},
  \bibinfo{pages}{249} (\bibinfo{year}{1954}).

\bibitem[{\citenamefont{Trickey et~al.}(1972)\citenamefont{Trickey, Kirk, and
  Adams}}]{tric72}
\bibinfo{author}{\bibfnamefont{S.~B.} \bibnamefont{Trickey}},
  \bibinfo{author}{\bibfnamefont{W.~P.} \bibnamefont{Kirk}}, \bibnamefont{and}
  \bibinfo{author}{\bibfnamefont{E.~D.} \bibnamefont{Adams}},
  \bibinfo{journal}{Rev.\ Mod.\ Phys.} \textbf{\bibinfo{volume}{44}},
  \bibinfo{pages}{668} (\bibinfo{year}{1972}).

\bibitem[{\citenamefont{Bakalov}(2012)}]{baka12}
\bibinfo{author}{\bibfnamefont{D.}~\bibnamefont{Bakalov}},
  \bibinfo{journal}{Hyperfine~Interact.} \textbf{\bibinfo{volume}{209}},
  \bibinfo{eid}{25} (\bibinfo{year}{2012}).

\bibitem[{\citenamefont{Clark and Ceperley}(2008)}]{clar08}
\bibinfo{author}{\bibfnamefont{B.~K.} \bibnamefont{Clark}} \bibnamefont{and}
  \bibinfo{author}{\bibfnamefont{D.~M.} \bibnamefont{Ceperley}},
  \bibinfo{journal}{Comput.\ Phys.\ Commun.} \textbf{\bibinfo{volume}{179}},
  \bibinfo{pages}{82} (\bibinfo{year}{2008}).

\bibitem[{\citenamefont{Whitlock et~al.}(1979)\citenamefont{Whitlock, Ceperley,
  Chester, and Kalos}}]{whit79}
\bibinfo{author}{\bibfnamefont{P.~A.} \bibnamefont{Whitlock}},
  \bibinfo{author}{\bibfnamefont{D.~M.} \bibnamefont{Ceperley}},
  \bibinfo{author}{\bibfnamefont{G.~V.} \bibnamefont{Chester}},
  \bibnamefont{and} \bibinfo{author}{\bibfnamefont{M.~H.} \bibnamefont{Kalos}},
  \bibinfo{journal}{Phys.\ Rev.~B} \textbf{\bibinfo{volume}{19}},
  \bibinfo{pages}{5598} (\bibinfo{year}{1979}).

\bibitem[{\citenamefont{Sears et~al.}(1979)\citenamefont{Sears, Svensson,
  Woods, and Martel}}]{sear79}
\bibinfo{author}{\bibfnamefont{V.~F.} \bibnamefont{Sears}},
  \bibinfo{author}{\bibfnamefont{E.~C.} \bibnamefont{Svensson}},
  \bibinfo{author}{\bibfnamefont{A.~D.~B.} \bibnamefont{Woods}},
  \bibnamefont{and} \bibinfo{author}{\bibfnamefont{P.}~\bibnamefont{Martel}},
  \bibinfo{type}{Tech. Rep.} \bibinfo{number}{AECL-6779},
  \bibinfo{institution}{{Atomic Energy of Canada Limited}}
  (\bibinfo{year}{1979}).

\end{thebibliography}

\end{document}